# Applications of wavelet transform in classification of local field potential recorded from the rat brain in conditioned place preference paradigm

Amirali Kalbasi [a], Mahdi Aliyari Shoorehdeli [a, *], Shole Jamali [b], Abbas Haghparast [c]

[a] *Department of Mechatronics, Faculty of Electrical Engineering, K. N. Toosi University of Technology, Tehran, Iran*

[b] *Department of Neuroscience, Medical University of South Carolina, Charleston, SC, USA*

[c] *Neuroscience Research Center, School of Medicine, Shahid Beheshti University of Medical Sciences, Tehran, Iran*

**Short Abstract:**

This study investigates the multi-label classification of Local Field Potential (LFP) data from the hippocampus (HIP) and nucleus accumbens (NAc) in the rat brain, focusing on reward responses using the Conditioned Place Preference (CPP) paradigm. Rats were conditioned with saline, morphine, and food rewards, and LFP recordings were conducted from both HIP and NAc during pre- and post-tests. The LFP data were classified into four categories: treatment types, test phases, recording channels, and chamber positions within the CPP setup. Features were extracted using Continuous Wavelet Transform (CWT), Wavelet Coherence, and Wavelet Scattering. Classification was performed via Decision Trees, Multilayer Perceptrons, and Support Vector Machines. Notably, in the Food group, HIP and combined HIP-NAc features yielded the highest classification accuracy for CPP chambers, whereas NAc features excelled in the Morphine group. Employing wavelet scattering, an 80% classification accuracy was achieved across treatment groups, test phases, and channels. Exceptionally high classification accuracies were observed for Food-post-test-HIP (99.75%) and Morphine-post-test-NAc (99.58%). The study reveals that NAc activity is pivotal for morphine-induced CPP, whereas HIP and HIP-NAc connectivity are crucial for food-induced CPP. The proposed methodology provides a novel avenue for precisely classifying LFP data, shedding light on neural circuit activities underlying behavioral responses.

**Abstract**

**Background and objective** Time-series classification is a critical component in data analysis, with wavelets serving as a powerful tool for both denoising and classifying biomedical data. Local Field Potential (LFP) recordings provide valuable insights into brain neural network activities, advancing our comprehension of the neural circuitry underlying various behaviors. This study aims to explore the multi-label classification of LFP data recorded from the hippocampus (HIP) and nucleus accumbens (NAc) of rat brain, in response to different reward stimuli.

**Methods** Rats were conditioned to saline, morphine, and food as rewards in the conditioned place preference (CPP) paradigm. LFP recordings were performed simultaneously from HIP and NAc in freely moving rats in pre- and post-test of CPP. LFP data were segmented into four categories: treatment groups (saline, morphine, and food), test phases (pre- and post-test), recording channels (HIP and NAc), and position of the animal in CPP chambers. Continuous wavelet transform (CWT) and wavelet coherence were used to extract features from the LFP signal. Then, data were classified to CPP chambers using decision trees and multilayer perceptron. Also, wavelet scattering was used to extract features from the data, and support vector machines (SVM) were used to classify data into treatment groups, test phases, and recording channels.

**Results** The study reveals that in the Food group, HIP and HIP-NAc features yielded the highest classification accuracy for CPP chambers, whereas NAc features were most accurate in the Morphine group. Remarkably, classifications using wavelet scattering and one-versus-all SVM achieved an 80% accuracy rate across different groups, test phases, and recording channels. Notably, the Food group's post-test HIP and Morphine group's post-test NAc exhibited the highest accuracy among all classification paradigms, at 99.75% and 99.58%, respectively.

**Conclusion** The findings demonstrate the pivotal role of NAc activity in morphine-induced CPP and the critical importance of HIP and HIP-NAc connectivity in food-induced CPP. The methodologies introduced in this study provide a robust framework for the accurate classification of LFP data, enhancing our understanding of neural circuit activities that drive behavior.

## 1. Introduction

Data classification is crucial in data analysis, leveraging both linear and nonlinear techniques. Linear classifiers like Logistic Regression [1], multilayer perceptron (MLP) [2], and support vector machine (SVM) [3] use a linear combination of input features to classify data points. Recent trends emphasize nonlinear and deep learning methods for enhanced classification performance [4].

The success and efficiency of these techniques heavily depend on the quality of feature extraction, achievable through time, frequency, and time-frequency domains. The time-frequency domain is particularly advantageous, capturing simultaneous time and frequency variations to provide more comprehensive information [5].

Feature extraction and classification of neural data from humans and animals represent a pivotal yet complex undertaking with multiple applications. These range from disease diagnosis [6], assessing the effect of different treatments [7] to conducting specialized studies [8]. Wavelet transforms stand as a key technique for transitioning data from the time domain to the time-frequency domain. Over the past decade, wavelets have proven to be invaluable tools for both denoising and classifying biomedical data [9-11].

In this study, local field potential (LFP) was recorded simultaneously from the nucleus accumbens (NAc) and hippocampus (HIP) of the rat brain as two important areas in the reward circuit to investigate the activity of these areas following natural and drug reward. Rats received food (natural reward), morphine (drug reward), and saline (control) during the conditioning phase of the conditioned place preference (CPP) task.

The purpose of the present study was the multi-label classification of the LFP recorded data into:

I. Treatment groups (food, morphine, or saline) administered during the conditioning phase.
II. Recording channels originating from either the HIP or NAc.
III. Position of the rat in CPP chambers.
IV. Pre- and post- conditioning phase (test phase: pre- and post-test)

In order to classify data, features were extracted from LFP data using I. Continuous wavelet transforms (CWT) method to transform data from the time-domain to the time-frequency domain, II. Wavelet scattering to obtain low-variance features from signals for machine learning and deep learning applications, as well as to automatically obtain features that minimize the difference between classes while preserving discriminability between them, and III. Wavelet coherence to measure and extract the correlation between HIP and NAc in the time-frequency domain. Also, various classification methods, including MLP, Decision trees, and SVM, were used to classify data into the desired classes.

## 2. Data recording and structure

### 2.1. Animals and Surgery

Male Wistar rats (Pasteur Institute, Tehran, Iran) weighing 220–270 g at the start of the experiment were maintained on a standard condition (12/12 h light/dark cycle in temperature (25 ± 2 °C) and humidity (55±10%). Animals were food restricted to 80-85% of their free-feeding body weight [12]. The Ethics Committee approved all Shahid Beheshti University of Medical Sciences (IR.SBMU.SM.REC.1395.373), Tehran, Iran, and followed the NIH standard (NIH publication No. 80-23 revised in 1996). The animals were anesthetized [13] and implanted with the recording electrodes to record LFPs in the NAc and hippocampal CA1 (HIP) [14] and at the following coordinates, respectively: anteroposterior (AP): -3.4 mm from bregma, lateral (L): ±2.5 mm, dorsal-ventral (DV): -2.6 mm and for NAc: AP: 1.5 mm, L: ±1.5 mm, DV: -7.6 mm. The reference and ground screws were inserted into the skull. The rats were allowed to recover for one week following surgery. At the end of the experiments, the electrode tip traces were localized and confirmed using a rat brain atlas (Paxinos and Watson 2007) [15].

*2.2. Conditioned place preference paradigm*

All rats experienced an unbiased CPP. LFP recordings were performed simultaneously from NAc and HIP in freely moving rats in pre- and post-test CPP. The CPP includes three phases: pre-conditioning, conditioning, and post-conditioning. The CPP apparatus consisted of two equal-sized compartments as the main chambers for conditioning reward and a smaller chamber (Null) connecting the two main chambers. One of the compartments' walls was striped vertically, and the other compartment had a horizontally-striped wall. The floor texture (smooth or rough) and wall strip pattern made the two main different compartments.

Behavior was monitored through a 3CCD camera (Panasonic, Japan) positioned above the apparatus. Data were analyzed by Ethovision software (Noldus Information Technology, the Netherlands), a video tracking system for automating behavioral experiments that were programmed to simultaneously trigger the onset of behavioral tracking and the beginning of LFPs recording. Behavior and electrophysiological sessions were recorded in a sync manner [16].

*2.3. Pre-conditioning phase (Pre-test)*

Before the conditioning phase, rats were examined for a 10-min pre-test (Fig. 1. A top panel) in which they had access to the entire CPP arena; the behavioral and LFP signals were recorded during this session. The CPP scores were calculated as the time spent in the rewarded compartment minus the unrewarded compartment's time. The total distance traveled (cm) was considered as the locomotor activity index for each animal.

*2.4. Conditioning phase (Saline, Morphine, Food)*

On the first day of the conditioning phase (Fig. 1. A middle panel), each animal received morphine (5 mg/kg, s.c.) in the morning and was confined to one chamber for 30 min; about six h later, the animals were injected with saline as the vehicle (1 ml/kg, s.c.) and were confined to another main chamber of CPP compartment for 30 min. On an alternate day, morphine and saline injections time were arranged in a counterbalanced manner. The third day of conditioning was the same as the first day. During this phase, access to other chambers of the CPP box was blocked. In the natural (food) group, on the first day of the conditioning period, in the morning session, food-restricted animals received 6 g biscuit as a reward in the middle of the one main compartment, and six hours later, they were placed into the other compartment with no food; each session lasted 30 min. On the following days, biscuit and no-food session times were arranged in a counterbalanced manner over the conditioning period. Throughout the experiment, animals were maintained on a restricted diet at 80-85% of their free-feeding weight but had access to water ad libitum at all times. As a control group in the saline group, animals just received saline in either the main compartment [16, 17]

*2.5. Post-conditioning phase (Post-test)*

Twenty-four hours following the conditioning phase, rats were examined for a 10-min post-test (Fig. 1. A bottom panel) in which they had access to the entire CPP arena, similar to the pre-test; the behavioral and LFP signals were recorded during this session. The CPP scores were calculated as the time spent in the rewarded compartment minus the unrewarded compartment's time. The total distance traveled (cm) was considered as the locomotor activity index for each animal.

During the pre-and post-test phases, animals freely explored the entire arena for 10 minutes while they were connected to the recording cable [17].

*2.6. Behavioral and electrophysiological recordings*

A digital video camera was used to record behavioral data (30 frames per second), while electrophysiological recordings and behavioral data were synchronized by a system that tracked the rat's movements. Each frame defined a spatial position as the center of the animal's body. During the experiments, the head-stage preamplifier pins were connected to a lightweight and flexible cable. A commercial acquisition processor (Niktek, IR) was used to record, digitize, and filter neural activity.

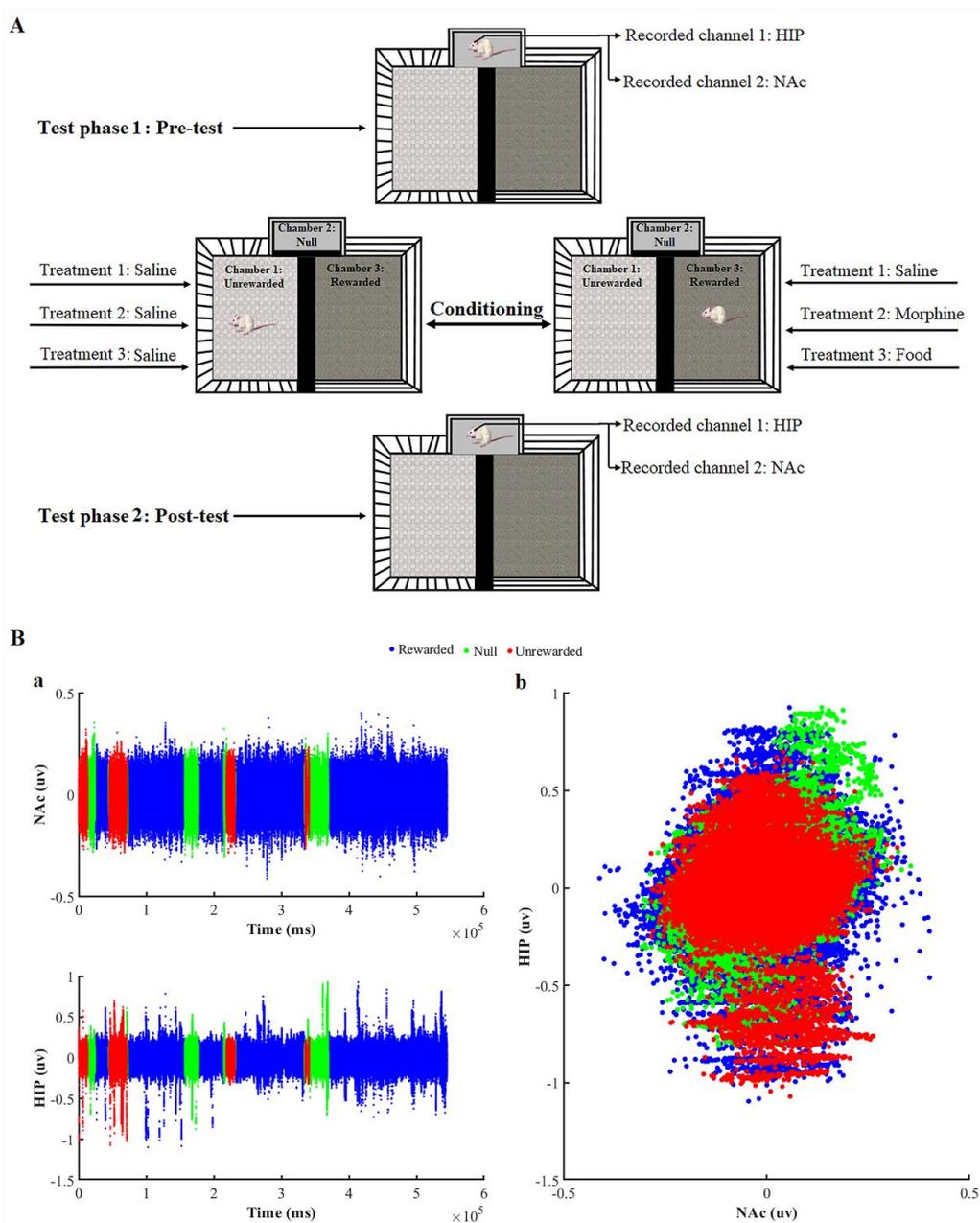

**Fig. 1. An overview of the behavioral and LFP recording structure: A.** Conditioned place preference paradigm (CPP), top panel: pre-test; middle panel: conditioning phase in which rats received saline, morphine, or food; bottom panel: post-test. Data labeled into four categories: I) Treatment groups (saline, morphine, food), II) Test phase (pre-and post-test), III) Recording Channels (HIP, NAc), and IV) Chambers (rewarded, null, and unrewarded). **B.** Raw data. B.a. top panel: NAc raw data; bottom panel: HIP raw data; b. HIP-NAc data; blue dots: rewarded, red dots: null, and green dots: unrewarded chambers.

*2.7. Data structure*

Data labeled in four categories: A) Treatment-groups: 1. saline group consisted of seven rats, 2. morphine group consisted of six rats, and 3. food group consisted of six rats (fig.1.A, middle panel); B) Testing phase: 1. pre-test and 2. post-test (fig.1.A, top and bottom panels); C) Chambers: 1. unrewarded, 2. null and 3. rewarded; D) Recording channels: 1. HIP and 2. NAc (fig. 1. B) .

## 3. Methodology

Fig.1. B.b shows that recorded data from HIP and NAc in different CPP chambers is not linearly separable. This suggests that wavelet-based time-frequency features could potentially distinguish between chambers based on the data derived from the HIP and NAc. For data classification, we employed multiple feature extraction techniques on the Local Field Potentials (LFP) data. First, Continuous Wavelet Transforms (CWT) were utilized to transition the data from the time domain to the time-frequency domain. Second, wavelet scattering was applied to secure low-variance features, optimally suited for machine learning and deep learning applications. This technique also served to automatically identify features that reduced inter-class variance while enhancing discriminability. Third, wavelet coherence analysis was conducted to quantify and capture correlations between the activities of the Hippocampus (HIP) and the Nucleus Accumbens (NAc) in the time-frequency domain. To execute the classification task, a range of algorithms, including Multilayer Perceptrons (MLP), Decision Trees, and Support Vector Machines (SVM), were deployed.

### 3.1. Continuous wavelet transform

LFP data was transformed from the time domain to the time-frequency domain using the wavelet transform, which is a variable-length window analysis. Initially, it examines the signal from a large scale or window to extract the most important features. The next step examines the signal with small windows and extracts the small characteristics of the signal. The continuous wavelet transform (CWT) is widely utilized in analyzing neural signals due to its adaptability in capturing transient features [18]. CWT is defined in (1).

$$C(scale, time) = \int_{-\infty}^{\infty} f(t) \times \left(\frac{1}{\sqrt{scale}}\right) \times \Psi\left(\frac{t-time}{scale}\right) dt \qquad (1)$$

In (1), "f" is an input signal. "Ψ" is a wavelet mother signal. "time" represents the wavelet function shifting, and "scale" scales the wavelet function's length, either dilating or compressing a signal [19]. We used the Morse wavelet mother function (**Error! Reference source not found.**, which has been demonstrated to provide optimal time-frequency localization in neural signal analysis [20]. A. The Fourier transform of the generalized Morse wavelet is as follows:

$$\Psi_{P,\gamma}(\omega) = U(\omega) a_{P,\gamma} \omega^{\frac{p^2}{\gamma}} e^{-\omega^\gamma} \qquad (2)$$

Where $U(\omega)$ is the unit step, $a_{P,\gamma}$ is a normalizing constant, $p^2$ is the time-bandwidth product, and $\gamma$ denotes the Morse wavelet's symmetry. Data can classify using the CWT's outputs, extracted features, and images [21]. CWT of Rat1 from the food group is shown in fig2. B.

## 3.2. Wavelet coherence

Wavelet coherence (WCOH) serves to ascertain the nonlinear interdependencies between two signals within the time-frequency domain [22]. Meanwhile, the wavelet cross-spectrum quantifies the power distribution shared between these signals, offering insights into their relationship across different frequency scales [23].

The wavelet cross-spectrum of two-time series, x, and y, is calculated as (3), where Cx(a,b) and Cy(a,b) correspond to continuous wavelet transforms of x and y with "a" as the scaling factor and "b" as the shifting factor. The superscript "*" indicates the complex conjugate, and "S" is a smoothing operator in time and scale [24].

$$C_{xy}(a,b) = S\left(C_x^*(a,b)C_y(a,b)\right) \qquad (3)$$

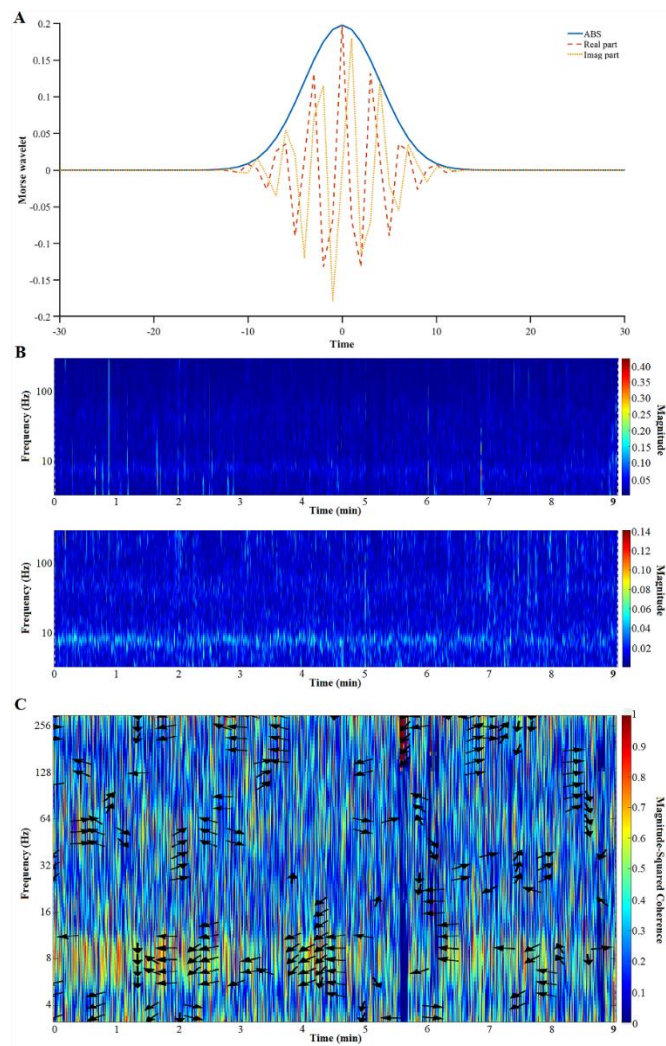

**Fig. 2. Feature extraction: A.** Generalized Morse mother wavelet, blue line: absolute value (ABS), red dashed line: real part, orange dotted line: imaginary part; **B.** Wavelet-transformed of LFP recorded from rat brain in the food group, top panel HIP CWT, bottom panel: NAc CWT; **C.** HIP-NAc WCOH of LFP recorded from rat brain in the food group. The phase lag between HIP and NAc is displayed with arrows on plots that use sampling frequency for areas where coherence exceeds 0.5. Arrows are spaced in time and scale to give a visual representation of the phase lag. The directions of the arrows correspond to the unit circle's phase lags. For example, the vertical arrow indicates a 1/2 or quarter cycle phase lag.

The wavelet coherence of two-time series x and y is calculated as (4) [25], and the wavelet coherence of Rat1 of the food group is shown in Fig. 2. C".

$$WCOH_{xy} = |C_{xy}(a,b)|^2 = |S(C_x^*(a,b)C_y(a,b))|^2 = S(|C_x(a,b)|^2) \cdot S(|C_y(a,b)|^2) \qquad (3)$$

*3.3. Multilayer perceptron*

A Multilayer Perceptron (MLP) is a type of supervised, feed-forward neural network composed of three primary components: an input layer, one or more hidden layers, and an output layer. Notably, all nodes—excluding those in the input layer—utilize a nonlinear activation function [26].

*3.4. Convolutional neural network*

Convolutional neural networks (CNNs) are artificial neural networks with three main components: A) The convolution layer filters the input data and extracts features from the data. B) A nonlinear operator, usually the ReLU function, applies as an activation function. C) Pooling for reducing the size of data. Unlike MLP, CNN determines the spatial relation of input. CNN is widely used for image feature extraction and classification [27, 28].

*3.5. Wavelet scattering network*

The scattering neural network consists of a convolutional neural network that has a fixed weight. The convolutional network filters the data first, then applies nonlinearity, and then pools or averages the results. The use of deep CNNs is associated with several challenges, including the need for large datasets and significant computing resources for training and evaluation, as well as the difficulty in understanding and interpreting the extracted features [29, 30].

Wavelet scattering addresses these challenges by employing a set of well-known filters and consists of three layers. In layer zero, a wavelet low-pass filter is applied to the input signal in order to average it. To incorporate the high-frequency information lost in layer 0, CWT is applied to the data at layer one to produce a set of scalogram coefficients [31]. Layer one filters the output by applying a nonlinear operator to the scalogram coefficients, which are then filtered by a wavelet low-pass filter. The output of layer one is considered as input for layer two. A wavelet low pass filter is applied to layer two in the same manner as layer zero, and the output is filtered. Several scattering coefficient layers can be used, but the energy dissipates with each iteration, so three layers are usually sufficient for most applications "Fig. " [32].

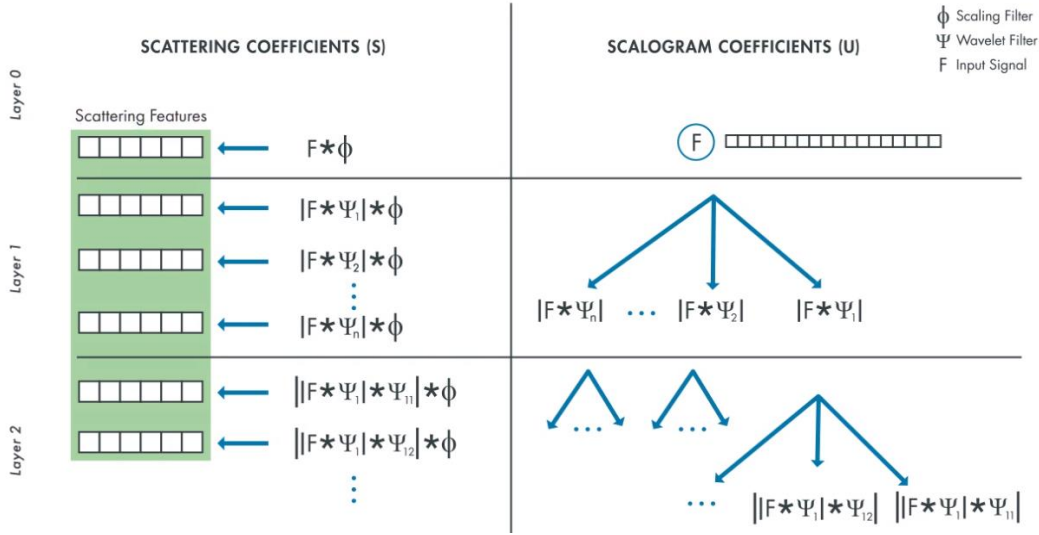

**Fig. 3. Wavelet scattering [33]:** The input signal is averaged in layer zero using a wavelet low-pass filter. In layer one, a continuous wavelet transform is applied to the data to generate a scalogram coefficient set that incorporates the high-frequency information lost in layer zero. Layer one filters the output with a wavelet low-pass filter after applying a nonlinear operator to scalogram coefficients. Layer one's scalogram coefficient is used as an input in layer two. Wavelet low pass filtering was applied to layer two as well as layer zero.

*3.6. Decision tree*

A decision tree is a type of supervised learning method used for classification as well as regression, which approximates a set of if-then-else decision rules based on data. In a decision tree, the classification or regression is modeled as a tree structure, in which the decision tree is incrementally constructed by subdividing the data set into smaller features [34].

*3.7. Support vector machine*

SVM is a non-parametric statistical method of supervised learning that is used for classification and regression. The SVM classifier relies on linear classification. It attempts to select the line (hyperplane) with the lowest error margin. Essentially, the support vector machine calculates the distance between the nearest data sample and the separator line (the boundary between categories) [35].

**4. Results**

This study aimed to achieve diverse classifications by leveraging wavelet-derived features. The classification schema encompassed four primary categories:

A) Treatment groups (Saline, Morphine, Food), B) Test phase (Pre-test, Post-test), C) Recording channels (HIP and NAc), D) Chambers (Rewarded, Null, Unrewarded).

"Fig. 4.A" delineates the six methodologies employed for chamber classification:
 I) HIP CWT coupled with MLP, II) NAc CWT coupled with MLP, III) HIP CWT coupled with DT, IV) NAc CWT coupled with DT, V) HIP-NAc WCOH coupled with MLP and VI) HIP-NAc WCOH coupled with DT.

The objective was to ascertain which channel (HIP, NAc, or the combined HIP-NAc) and which classification algorithm (MLP or DT) delivered the most effective chamber categorization. Decision Tree complexity metrics were also calculated and are presented in Figure 5. Additionally, Figure 4.B illustrates the application of an alternative method—combining wavelet scattering with Support Vector Machine (SVM) classifiers—to segregate data into treatment groups, test phases, and recording channels.

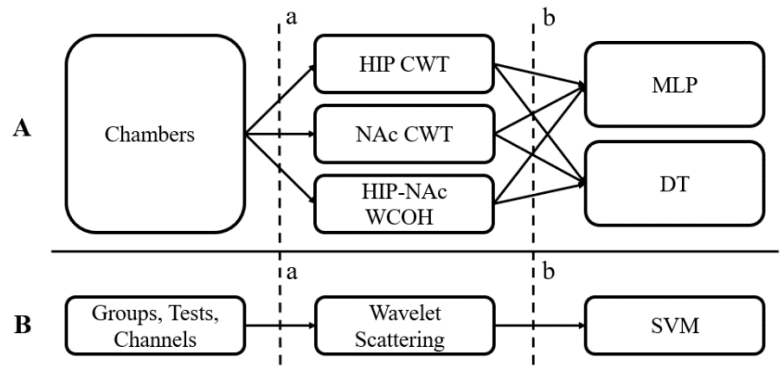

**Fig. 4. Classification structure. A.** Methods to classify the condition of rat in CPP box (Rewarded, Null, Unrewarded), Aa Feature extraction methods including HIP CWT, NAc CWT, and HIP-NAc WCOH, Ab Classification methods including decision tree and MLP; **B.** Methods to classify groups (Saline, Morphine, and Food), tests (Pre- and Post- tests) and channels (HIP and NAc), **B.a.** Feature extraction methods including wavelet scattering and CWT, **B.b.** SVM for classification.

TABLE 1: DT RESULTS

|  | Food | Morphine | Saline |
|---|---|---|---|
| HIP | 100 | 96 | 98 |
| NAc | 97 | 100 | 98 |
| HIP-NAc | 100 | 92 | 96 |

TABLE 2: DT COMPLEXITY

|  | Food | Morphine | Saline |
|---|---|---|---|
| HIP | Low | Mid | Mid |
| NAc | High | Low | Mid |
| HIP-NAc | Low | High | High |

TABLE 3: MLP RESULTS

|  | Food | Morphine | Saline |
|---|---|---|---|
| HIP | 98 | 93 | 95 |
| NAc | 94 | 97 | 94 |
| HIP-NAc | 97 | 88 | 91 |

*4.1. Conditions classification*

The average accuracy metrics for classifications using HIP or NAc Continuous Wavelet Transform (CWT) coupled with Decision Trees (DT), as well as HIP-NAc Wavelet Coherence (WCOH) paired with DT, are presented in Table 1. Corresponding Decision Tree complexity for these methods is outlined in Table 2. Additionally, Table 3 provides the average accuracy for classifications employing HIP or NAc CWT in conjunction with Multilayer Perceptrons (MLP), and HIP-NAc WCOH combined with MLP.

As indicated in Table 1, the Food group exhibited optimal classification accuracy when utilizing either HIP (100%) or HIP-NAc (100%) features for chamber identification. In contrast, the Morphine group achieved peak accuracy with NAc features (100%), while the Saline group showed no preferential accuracy between HIP (98%) and NAc (98%). Notably, HIP-NAc classification in the Morphine group underperformed, yielding only a 92% accuracy rate. When utilizing Wavelet Coherence (WCOH) with a Decision Tree (DT) classifier, the Food group (100%) outperformed both the Saline (96%) and Morphine (92%) groups.

Table 2 reveals that the Food group exhibited lower DT complexity when employing HIP Continuous Wavelet Transform (CWT) and HIP-NAc WCOH, whereas the lowest DT complexity in the Morphine group was observed with NAc CWT.

According to Table 3, MLP classification results aligned with those from DT but with slightly diminished performance. Specifically, NAc CWT combined with MLP yielded the highest accuracy in the Morphine group (97%). In the Food group, both HIP CWT and HIP-NAc WCOH, when coupled with MLP, achieved top performance with 98% and 97% accuracy, respectively. In the Saline group, no significant difference was noted between HIP (95%) and NAc (94%) when paired with MLP. Moreover, HIP-NAc WCOH coupled with MLP in the Food group (97%) surpassed the accuracies in both the Saline (91%) and Morphine (88%) groups.

These findings demonstrate that Decision Trees (DT) outperformed Multilayer Perceptrons (MLP) in classifying Conditioned Place Preference (CPP) chambers based on Local Field Potential (LFP) data from the Hippocampus (HIP) and Nucleus Accumbens (NAc). As Tables 1-3 suggest, in the Food group, both HIP Continuous Wavelet Transform (CWT) and HIP-NAc Wavelet Coherence (WCOH) yielded superior results with reduced model complexity. Conversely, in the Morphine group, NAc features stood out in terms of both performance and lower complexity. The Saline group showed no discernible difference between HIP and NAc in terms of CWT performance or model complexity. Additionally, HIP-NAc WCOH in the Food group exhibited both the highest accuracy and the least model complexity when compared to the Saline and Morphine groups. These observations imply that HIP and HIP-NAc connectivity play a critical role in CPP chamber classification within the Food group, while NAc activity takes precedence in the Morphine group

*4.2. Groups, channels, and test classification*

We employed wavelet scattering to feature-extract from the Local Field Potential (LFP) data recorded from the Hippocampus (HIP) and Nucleus Accumbens (NAc). The data was subsequently classified using a one-versus-all Support Vector Machine (SVM) approach, validated through K-fold cross-validation with K=10 (See Figure 6). Intriguingly, this methodology achieved a simultaneous classification of groups, tests, and channels in the rat subjects with an 80% accuracy rate.

**Fig. 5. Decision-tree (DT) complexity sample for a rat in food group:** A. DT results for the CWT of the HIP channel; B. DT results for a HIP-NAc WCOH.

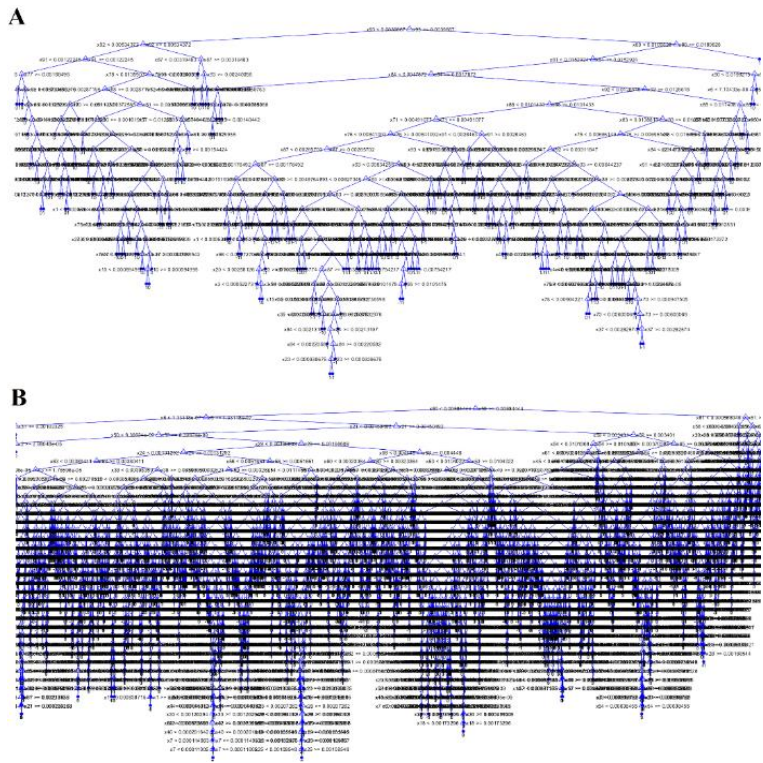

| True Class | Predicted Class | | | | | | | | | | | | True positive rates (TPR) | False negative rate (FNR) |
|---|---|---|---|---|---|---|---|---|---|---|---|---|---|---|
| HIP-Post-Morphine | 1773 | 2 | 209 | 143 | 115 | 99 | 1 | 1 | 1 | 1 | 1 | 0 | 75.57544757 | 24.42455243 |
| HIP-Post-Food | 0 | 2354 | 0 | 0 | 0 | 1 | 1 | 0 | 1 | 2 | 1 | 0 | 99.74576271 | 0.254237288 |
| HIP-Post-Saline | 158 | 2 | 2091 | 150 | 231 | 95 | 2 | 1 | 2 | 0 | 3 | 2 | 76.39751553 | 23.60248447 |
| HIP-Pre-Morphine | 143 | 1 | 96 | 1805 | 135 | 160 | 1 | 0 | 1 | 1 | 1 | 2 | 76.93947144 | 23.06052856 |
| HIP-Pre-Food | 99 | 2 | 139 | 162 | 1697 | 141 | 0 | 1 | 2 | 1 | 1 | 1 | 75.55654497 | 24.44345503 |
| HIP-Pre-Saline | 124 | 8 | 106 | 118 | 318 | 2106 | 1 | 2 | 3 | 2 | 1 | 0 | 75.51093582 | 24.48906418 |
| NAc-Post-Morphine | 1 | 0 | 0 | 0 | 1 | 2 | 2343 | 1 | 2 | 2 | 0 | 1 | 99.57501062 | 0.424989375 |
| NAc-Post-Food | 0 | 2 | 3 | 2 | 1 | 2 | 6 | 1765 | 182 | 40 | 283 | 60 | 75.2344416 | 24.7655584 |
| NAc-Post-Saline | 1 | 4 | 1 | 1 | 2 | 2 | 5 | 180 | 2113 | 68 | 70 | 290 | 77.20131531 | 22.79868469 |
| NAc-Pre-Morphine | 1 | 2 | 0 | 2 | 3 | 2 | 3 | 114 | 142 | 1768 | 136 | 173 | 75.36231884 | 24.63768116 |
| NAc-Pre-Food | 2 | 3 | 3 | 1 | 1 | 2 | 1 | 180 | 80 | 114 | 1827 | 132 | 77.87723785 | 22.12276215 |
| NAc-Pre-Saline | 1 | 1 | 0 | 2 | 1 | 1 | 2 | 105 | 220 | 167 | 182 | 2055 | 75.0822068 | 24.9177932 |
| | HIP-Post-Morphine | HIP-Post-Food | HIP-Post-Saline | HIP-Pre-Morphine | HIP-Pre-Food | HIP-Pre-Saline | NAc-Post-Morphine | NAc-Post-Food | NAc-Post-Saline | NAc-Pre-Morphine | NAc-Pre-Food | NAc-Pre-Saline | Total Accuarcy | 80.00485076 |

**Fig. 6. Confusion chart of wavelet scattering coupled with SVM classifier with k-fold cross-validation.** Left panel: K-fold classification results (K=10), right panel: true positive rate (TPR) refers to the probability of a predicted class, conditioned on truly labeling, false negative rate (FNR) refers to the probability of a predicted class, conditioned on falsely labeling, and total accuracy of this method.

Further analysis revealed that the LFP data from the HIP channel during the post-test in the Food treatment group exhibited exceptional classification performance at 99.75% (See Figure 5, right panel, dark blue). Similarly, the LFP data from the NAc channel during the post-test in the Morphine treatment group yielded a 99.58% accuracy (See Figure 5, right panel, dark blue). These results significantly outperformed the classification of LFP data from the HIP in the Food group (75.56%) and NAc in the Morphine group (75.36%) during the pre-test phase. Given that the rats were not conditioned prior to the pre-test, it was expected that their signal characteristics would be relatively uniform before conditioning.

## 5. Conclusion

The aim of this study was to employ wavelet-derived features for classifying Local Field Potential (LFP) data into various categories: chambers, groups, tests, and channels. Our findings demonstrate that utilizing Continuous Wavelet Transform (CWT) features in conjunction with Decision Tree (DT) and Multilayer Perceptron (MLP) classifiers yielded the most effective chamber classification based on Hippocampus (HIP) data in the Food group. Conversely, Nucleus Accumbens (NAc) data proved most effective in the Morphine group.

Furthermore, combining HIP and NAc features—specifically through HIP-NAc Wavelet Coherence (WCOH)—resulted in the highest accuracy rates when using both MLP and DT classifiers in the Food group. Additionally, classifications for groups, channels, and tests, using features extracted via wavelet scattering and coupled with a Support Vector Machine (SVM) classifier, aligned well with the chamber classifications. This suggests that LFP data from HIP and NAc are particularly effective in the Food and Morphine groups, respectively.

Consequently, in the Morphine group, NAc-based LFP data could comprehensively classify the animal's group, test phase, recording channel, and chamber location. In contrast, in the Food group, such classification is predominantly reliant on HIP-based data.

This study adds to the growing body of literature emphasizing the importance of targeted feature extraction in neural data classification. The differential effectiveness of HIP and NAc in various groups underscores the need for a context-specific approach in feature selection and classification methodologies. Future work could explore the incorporation of additional neural features or the use of more complex machine learning algorithms to further refine classification accuracy. Moreover, longitudinal studies could offer insights into how these neural activities evolve over time, enhancing our understanding of the underlying biological mechanisms.

In summary, our work lays the groundwork for more nuanced, effective strategies in the classification of neural data, holding promise for applications ranging from medical diagnostics to understanding complex neural networks in various physiological conditions.